\documentclass[reprint,superscriptaddress,amsmath,amssymb,pra]{revtex4-1}
\usepackage{graphicx,color}
\usepackage{bm}
\usepackage{newtxtext, newtxmath}

\begin{document}
\title{Generation of pseudo-sunlight via quantum entangled photons and the interaction with molecules}

\author{Yuta Fujihashi}
\affiliation{Institute for Molecular Science, National Institutes of Natural Sciences, Okazaki 444-8585, Japan}
\author{Ryosuke Shimizu}
\affiliation{Department of Engineering Science, University of Electro-Communications, Chofu 182-8585, Japan}
\author{Akihito Ishizaki}
\affiliation{Institute for Molecular Science, National Institutes of Natural Sciences, Okazaki 444-8585, Japan}
\affiliation{School of Physical Sciences, Graduate University for Advanced Studies, Okazaki 444-8585, Japan}


\begin{abstract}
Light incident upon molecules trigger fundamental processes in diverse systems present in nature. However, under natural conditions, such as sunlight illumination, it is impossible to assign known times for photon arrival owing to continuous pumping, and therefore, the photo-induced processes cannot be easily investigated. In this work, we theoretically demonstrate that characteristics of sunlight photons such as photon number statistics and spectral distribution can be emulated through quantum entangled photon pair generated with the parametric down-conversion (PDC). We show that the average photon number of the sunlight in a specific frequency spectrum, e.g., the visible light, can be reconstructed by adjusting the PDC crystal length and pump frequency, and thereby molecular dynamics induced by the pseudo-sunlight can be investigated. The entanglement time, which is the hallmark of quantum entangled photons, can serve as a control knob to resolve the photon arrival times, enabling investigations on real-time dynamics triggered by the pseudo-sunlight photons.
\end{abstract}
\maketitle

\section{Introduction}
Giant strides in ultrashort laser pulse technology have opened up real-time observation of dynamical processes in complex physical, chemical, and biological systems. Under natural conditions, such as sunlight illumination, known times cannot be assigned for photon arrival owing to continuous pumping, and thus, photo-induced dynamical processes cannot be easily investigated. In time-resolved optical spectroscopy, however, investigations on dynamical processes can be conducted by synchronizing the initial excitations in the entire ensemble with the use of ultrashort pulsed laser and thereby amplifying the microscopic dynamics in a constructively interferential fashion. In this manner, time-resolved laser spectroscopy has provided detailed information and deeper insights into microscopic processes in complex molecular systems. Nevertheless, the relevance of the laser spectroscopic data regarding photosynthetic proteins was challenged and whether dynamics initiated by sunlight irradiation might be different from those detected with laser spectroscopy is still being debated \cite{Cheng:2009co,Mancal:2010kc,Ishizaki:2011cx,Brumer:2012ib,Fassioli:2012gd,Chenu:2015hi,Brumer:2018ka,Chan:2018em,Shatokhin:2018jq,muoz2019photon}. 
Although spectroscopic measurements may or may not demonstrate phenomena under sunlight illumination in the one-to-one correspondence, the debate inspired us to comprehend the occurrence of photoexcitation under natural irradiation.

Sunlight is considered as the radiation from the black-body with an effective temperature of approximately 5800\,K, and thus, the coherence time is extremely short. 
Furthermore, the photon number statistics obeys the Bose--Einstein distribution, whereas the coherent laser is characterized by the Poisson photon number statistics \cite{mandel1995optical}. A variety of schemes have been proposed to generate light that mimics sunlight, e.g., the solar simulator with Xenon arc lamp. Broadband incoherent light that emulate the temporal property of the thermal light has been also investigated, e.g., scattered laser beam from a rotating ground-glass disc \cite{Estes:1971if} and amplified spontaneous emission \cite{Turner:2013fq}. 
It was also presented that the thermal light could be well represented in terms of a statistical mixture of laser pulses, which gives a valid description of linear light-matter interactions \cite{Chenu:2015km,Chenu:2016cx}.
However, these schemes do not provide knobs to control light on an ultrafast timescale approximate to a few femtoseconds, which is relevant for energy/charge transfer during the primary steps of photosynthesis and isomerization reaction in the first steps of vision. Therefore, a scheme to control thermal light on an ultrashort timescale to unveil how photoexcitation by natural light and the subsequent dynamics proceed should be developed.

To tackle this issue, we take a constructive approach instead of a direction toward manipulating the actual thermal light. We examine quantum states of photons that reconstruct characteristics of the sunlight, specifically statistical properties such as photon number statistics and spectral distribution. In this work, such photon states are termed pseudo-sunlight. An advantage of the approach is that deductively obtained expressions of the quantum states enable us to investigate photo-induced dynamical processes in molecular systems with quantitative underpinnings and help us gain deeper insights into physical implications. To this end, we address quantum entangled photon pairs generated through the parametric down-conversion (PDC) in birefringent crystals \cite{mandel1995optical}. The state vector of the generated pairs yields a form of the geometric distribution as the photon number probability \cite{Barnett:1985jc,Yurke:1987hs,Gerry2005:introductory}. In addition, the two photons in a pair exhibit continuous frequency entanglement stemming from the conservation of energy and momentum. Consequently, the quantum state of the one in the pair is a mixed state in terms of frequency with some spectral distribution, when the state of the other is not fully measured. For these reasons, the entangled photon pairs are expected to reconstruct statistical characteristics of the sunlight.

\section{Frequency-entangled photons}

We consider the PDC process, where a pump photon with frequency $\omega_{\rm p}$ is split into two entangled photons, signal and idler photons with frequencies $\omega_1$ and $\omega_2$ such that $\omega_{\rm p} = \omega_1 + \omega_2$. 
Electric fields inside a one-dimensional nonlinear crystal whose length is $L$ are considered and the time-ordering effect during the PDC process is neglected. This approximation is relevant in describing the PDC in the low-gain regime \cite{Christ:2013fg}, and the state vector of the generated photons is obtained as $\lvert \psi_{\rm PDC} \rangle = \exp(-i \hat{H}_{\rm PDC}/\hbar) \lvert {\rm vac} \rangle$ with 
$  
 	\hat{H}_{\rm PDC}
    =
    \int d\omega_1 \int d\omega_2 \,
    f (\omega_1, \omega_2)
    \hat{a}_{\rm s}^\dagger(\omega_1)
    \hat{a}_{\rm i}^\dagger(\omega_2)
    +
    {\rm h.c.}
$,
where $\hat{a}_{\rm s}^\dagger(\omega)$ and $\hat{a}_{\rm i}^\dagger(\omega)$ denote the creation operators of the signal and idler photons, respectively \cite{Grice:1997ht}. 
The two-photon amplitude $f (\omega_1, \omega_2)$ is expressed as  
$f (\omega_1, \omega_2) = \hbar B \alpha_{\rm p} (\omega_1 + \omega_2)\,\Phi (\omega_1, \omega_2)$,
where $\alpha_{\rm p} (\omega)$ is the pump envelope function normalized as $\int d\omega\,\alpha_{\rm p} (\omega)=1$ and $\Phi (\omega_1, \omega_2) = {\rm sinc}[\Delta k(\omega_1, \omega_2) L / 2]$ is referred to as the phase-matching function, where $\Delta k(\omega_1, \omega_2)$ represents the momentum mismatch among the input and output photons.
All other constants such as the second-order susceptibility of the crystal and pump intensity are merged into the factor $B$, which corresponds to the conversion efficiency of the PDC \cite{Grice:1997ht}. 
Typically, $\Delta k(\omega_1, \omega_2)$ may be approximated linearly around the central frequencies of the generated beams, $\bar\omega_{\rm s}$ and $\bar\omega_{\rm i}$ as
$\Delta k(\omega_1, \omega_2) L =  (\omega_1 - \bar{\omega}_{\rm s}) T_{\rm s} + (\omega_2 - \bar{\omega}_{\rm i} ) T_{\rm i}$ with $T_\sigma = L/v_{\rm p} - L/v_\sigma$ \cite{Grice:1997ht,Rubin:1994theory}, where $v_{\rm p}$  and $v_{\sigma}$ are the group velocities of the pump laser and a generated beam at frequency $\bar\omega_\sigma$, respectively. The difference $T_{\rm e} = \lvert T_{\rm s}-T_{\rm i} \rvert$ is termed the entanglement time \cite{Saleh:1998vl,Dorfman:2016da}, which represents the maximal time delay between the arrival of the two entangled photons.

\begin{figure}
    \includegraphics{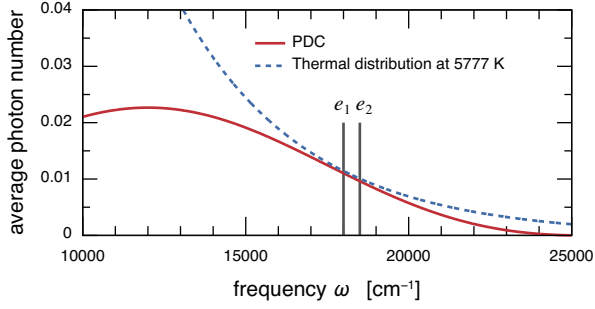}
    \caption{
	Average photon number in the signal beam generated through the PDC process with the CW pumping, $\bar{n}(\omega)$ in Eq.~\eqref{eq:photon-number-cw} (red solid line) and the thermal distribution, $\bar{n}_{\rm th}(\omega) = (e^{\hbar\omega/k_{\rm B}T}-1)^{-1}$ at temperature $T=5777\,{\rm K}$ (blue dashed line). To evaluate Eq.~\eqref{eq:photon-number-cw}, the parameters of $\omega_{\rm p}=25000\,{\rm cm}^{-1}$, $\bar{\omega}_{\rm s}=12000\,{\rm cm^{-1}}$, $T_{\rm e} = 2.5\,{\rm fs}$, and $B =0.15$ are employed. The two vertical lines indicate electronic transition energies that will be discussed later, $\omega_{1{\rm g}} = 18000\,{\rm cm^{-1}} $ and $\omega_{2{\rm g}} = 18500\,{\rm cm^{-1}} $.
    }
    \label{fig:1}
\end{figure}

For CW pumping, the two-photon amplitude
is written as $f (\omega_1,\omega_2) = \hbar \delta(\omega_1 + \omega_2 - \omega_{\rm p})\,r(\omega_1)$ with $ r(\omega) = B  \, \Phi (\omega,\omega_{\rm p} - \omega) = B\, {\rm sinc}[(\omega - \bar{\omega}_{\rm s}) T_{\rm e}/2]$. Therefore, the propagator is recast into $\exp(-i \hat{H}_{\rm PDC}/\hbar) = \exp[ \int\ d\omega\, r(\omega) \hat{a}_{\rm s}^\dagger(\omega) \hat{a}_{\rm i}^\dagger(\omega_{\rm p}-\omega) - {\rm h.c.} ]$, and the output photon state is obtained as the two-mode squeezed vacuum state \cite{Gerry2005:introductory},
\begin{align}
    \lvert \psi_{\rm PDC} \rangle
    =
    \prod_{\omega}
    \sum_{n_\omega=0}^\infty
    \frac{[\tanh r(\omega)]^{n_\omega}}{\cosh r(\omega)}
    \lvert n_\omega \rangle_{{\rm s}}
    \lvert n_{\omega_{\rm p} - \omega} \rangle_{{\rm i}},
    \label{eq:two-mode-squeeezed}
\end{align}
where $\lvert n_\omega \rangle_\sigma$ is the Fock state of the photon $\sigma$ with frequency $\omega$. When the idler photon is discarded without being measured, the quantum state of the signal photon is mixed \cite{Barnett:1985jc,Yurke:1987hs,Gerry2005:introductory}. This situation is described by tracing out the idler photon's degrees of freedom such that $\hat{\rho}_{\rm s} = {\rm tr}_{\rm i} ( \lvert \psi_{\rm PDC} \rangle \langle \psi_{\rm PDC} \rvert )$ \cite{mandel1995optical}. Consequently, the reduced density operator of the signal photon reads to $   
    \hat{\rho}_{\rm s}
    = 
    \prod_{\omega}
    \sum_{n_\omega =0}^\infty 
    P_\omega(n_\omega) 
    \lvert n_\omega \rangle_{\rm s} \langle n_\omega \rvert_{\rm s}
$, where $P_\omega(n)$ represents the probability that there are $n$ photons of frequency $\omega$ in the signal photon beam, $P_\omega(n) = [1-\zeta(\omega)] \zeta(\omega)^{n}$ with $\zeta(\omega) = \tanh^2 r(\omega)$. The density operator is also expressed as $\hat{\rho}_{\rm s} = Z^{-1} \exp[-\int d\omega \, \ln \zeta(\omega) \hat{a}_{\rm s}^\dagger(\omega) \hat{a}_{\rm s}(\omega)]$ with $Z$ being the partition function \cite{Quesada:2019gj}.
Therefore, the quantum state $\hat{\rho}_{\rm s}$ can be regarded as the thermal state in the sense that the photon-number statistics obey the geometric distribution, and the average photon number is computed as a function of $\omega$, 
\begin{align}
    \bar{n}(\omega)
    = 
    \sum_{n=0}^\infty n P_\omega(n)
    =
    \frac{\zeta(\omega)}{1-\zeta(\omega)}
    = 
    \sinh^2 r(\omega).
    \label{eq:photon-number-cw}
\end{align}
Specifically, when $\zeta(\omega)$ can be approximately expressed as $\zeta(\omega) \simeq \exp(-\hbar \omega /k_{\rm B}T)$, the expressions of $P_\omega(n)$ and $\bar{n}(\omega)$ become identical to those of the thermal radiation from a black-body with temperature $T$, where $k_{\rm B}$ denotes the Boltzmann constant. 
This is a necessary and sufficient condition for reconstructing statistical properties of the field and hence the field correlation functions at any order.
Although the whole frequency range may be impossible to be reconstructed, it is still beneficial to emulate a specific frequency region such as the visible light for unveiling how photoexcitation by natural light and the subsequent dynamics proceed, for example, in photosynthesis and vision.

Figure~\ref{fig:1} presents the average photon number in the signal beam generated through PDC with CW pumping. 
For comparison, the thermal distribution, $\bar{n}_{\rm th}(\omega) = (e^{\hbar\omega/k_{\rm B}T}-1)^{-1}$, at temperature $T=5777\,{\rm K}$ is also shown.
The parameters employed for the calculation are chosen so as to reproduce the average photon number of the visible light, as given in the figure caption. These values are realizable through the use of birefringent crystals such $\beta$-$\mathrm{BaB_2O_4}$
\cite{Grice:2001jc,Lee:2006id} and  $\mathrm{BiB_3O_6}$ \cite{maclean:2018direct}. 
Figure~\ref{fig:1} demonstrates that Eq.~\eqref{eq:photon-number-cw} is capable of approximately reproducing the average photon number of the black-body radiation in  the visible region, by adjusting the crystal length $L$ and pump frequency $\omega_{\rm p}$.
It is noteworthy that the photon number in the visible region is substantially 0 or 1.

\begin{figure}
    \includegraphics{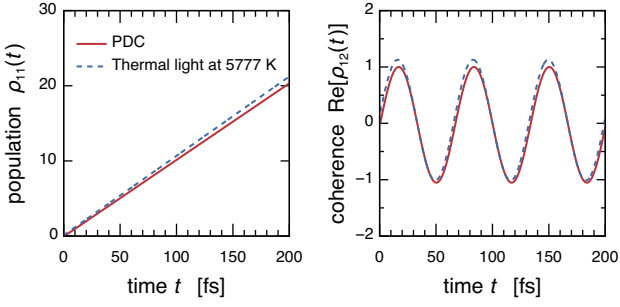}
    \caption{
	Time evolution of the normalized density matrix elements describing molecular electronic excitations, Eq.~\eqref{eq:unconditional-RDM1}, under illumination of the signal photons generated through the PDC process (red solid lines) and the black-body radiation of temperature 5777\,K (blue dashed line). The average photon number and the electronic transition energies shown in Fig.~\ref{fig:1} were used in the calculations. The normalization is such that the maximum value of the real part of the off-diagonal element, $\rho_{12}(t) = \langle e_1 \vert \hat\rho_{\rm el}(t) \vert e_2 \rangle$, for the PDC case is unity; therefore, it is of no consequence that the normalized population exceeds unity. 
    }
    \label{fig:2}
\end{figure}

\section{Interaction of signal photons with molecules}

Theoretical expressions of quantum states of the photons enable one to investigate photo-induced molecular dynamical processes with quantitative underpinnings. Here, we discuss the electronic excitation of a molecule. The molecule is modeled by the electronic ground state $\lvert g \rangle$ and electronic excited states $\{ \lvert e_\alpha \rangle\}_{\alpha=1,2,\dots}$, and the Hamiltonian is given by $ \hat{H}_{\rm mol} = \epsilon_g \lvert g \rangle \langle g \rvert + \sum_\alpha \epsilon_{\alpha} \lvert e_\alpha \rangle \langle e_\alpha \rvert$. The states $\{ \lvert e_\alpha \rangle\}$ correspond to electronic excitons in the single-excitation manifold of a molecular aggregate. The optical transitions between $\lvert g \rangle$ and $\lvert e_\alpha \rangle$ are described by the operator $\hat{\mu} = \sum_\alpha \mu_{\alpha g} ( \lvert e_\alpha \rangle \langle g \rvert + \lvert g \rangle \langle e_\alpha \rvert)$, where $\mu_{\alpha g}$ stands for the transition dipole. In general, environment-induced fluctuations in electronic energy strongly influence the excited-state dynamics in condensed phases. However, in this study we ignored the environmental degrees of freedom because the main concern here is to investigate the characteristics of pseudo-sunlight irradiation, which is qualitatively independent of the effects of the environment. For simplicity, radiative and nonradiative decays to the ground state are also neglected.

The following setup is considered: The signal and idler beams generated through PDC are split. Only signal photons interact with molecules, and idler photons propagate freely. Therefore, the total molecule-field Hamiltonian can be written as $\hat{H}_{\rm total} = \hat{H}_{\rm mol} + \hat{H}_{\rm field} + \hat{H}_{\rm int}$. The second term in this equation, $\hat{H}_{\rm field}$, is the free Hamiltonian of the signal and idler photons, and the molecule-field interaction is described by $\hat{H}_{\rm int}(t) = - \hat{\mu} \hat{E}_{\rm s}(t)$. Owing to the weak field-matter interaction, the first-order perturbative truncation in terms of $\hat{H}_{\rm int}(t)$ provides a reasonable description of the electronic excitation generated with the signal photon absorption.
Thus, the state vector to describe the molecular excitation together with signal and idler photons is obtained as  
$	\lvert \psi_{\rm total} (t) \rangle
	 =
    ({i}/{\hbar})
	\sum_{\alpha}
	\mu_{\alpha g}
	\int^t_{-\infty} d\tau_1\,
	e^{-i\omega_{\alpha g} (t-\tau_1)}
    \hat{E}_{\rm s}( \tau_1)
	\lvert e_\alpha  \rangle
	\lvert \psi_{\rm PDC}  \rangle$,
where $ \omega_{\alpha \beta} = ( \epsilon_\alpha - \epsilon_\beta)/\hbar$ has been introduced.
In the equation, the field operator of the signal photon $\hat{E}_{\rm s}(t)$ can be divided into positive- and negative-frequency components, $\hat{E}^{(+)}_{\rm s} (t) = \int d\omega\, i A (\omega )\, \hat{a}_{\rm s}(\omega) e^{-i\omega t}$ and $\hat{E}^{(-)}_{\rm s} (t)= [\hat{E}^{(+)}_{\rm s} (t)]^\dagger$, respectively, where $A (\omega )\propto\sqrt{\omega }$ \cite{loudon2000quantum}. The negative-frequency component causes the rapidly oscillating term in the integrand; hence, the contribution to the electronic excitation is negligibly small. Therefore, $\hat{E}_{\rm s}(\tau_1)$ can be replaced with the positive-frequency component. This rotating wave approximation is of no consequence in cases of weak field-matter interaction \cite{Chenu:2016cx,Dorfman:2016da}, but it breaks down in the strong interaction regime \cite{Zueco:2009it}.

When the quantum states of idler photons are not measured, the reduced density operator to describe the electronic excitation is obtained by tracing over the fields' degrees of freedom, $\hat{\rho}_{\rm el}(t) = {\rm tr}_{{\rm s}+{\rm i}} [ \lvert \psi_{\rm total} (t) \rangle \langle \psi_{\rm total} (t) \rvert ]$, as such:
\begin{align}
	\hat{\rho}_{\rm el}(t)
	&=
	\sum_{\alpha \beta}
	\frac{\mu_{\alpha g} \mu_{\beta g}}{\hbar^2}
	e^{-i\omega_{\alpha \beta} t}
	\int_{-\infty}^t d\tau_2\,
	e^{-i\omega_{\beta g} \tau_2}
\notag\\
    &\quad \times
	\int_{-\infty}^t d\tau_1\,
	e^{i\omega_{\alpha g} \tau_1}
	G_{\rm s}^{(1)}(\tau_2,\tau_1) 
	\lvert e_\alpha \rangle\langle e_\beta \rvert,
	\label{eq:unconditional-RDM1}
\end{align}
where the first-order temporal correlation function of signal photons $G_{\rm s}^{(1)}(t_2,t_1) = \langle \hat{E}_{\rm s}^{(-)}( t_2) \hat{E}_{\rm s}^{(+)}(t_1 ) \rangle$ has been introduced. In the CW pumping case, the function is expressed as \cite{Quesada:2019gj} 
\begin{align}
    G_{\rm s}^{(1)} (t_2,t_1)
    =
    \int^\infty_0 d\omega \,
    e^{i\omega (t_2-t_1)}
    A (\omega)^2
    \bar{n}(\omega).
    \label{eq:1st-Glauber-CW}
\end{align}
By replacing $\bar{n}(\omega)$ with the average photon number for thermal light $\bar{n}_{\rm th}(\omega) = ( e^{\hbar \omega/ k_{\rm B} T} -1 )^{-1}$, Eq.~\eqref{eq:unconditional-RDM1} becomes practically identical to the expression for the density operator under the influence of the real sunlight photons \cite{Brumer:2018ka}.
Figure~\ref{fig:2} demonstrates the time evolution of the matrix elements of the density operator under the illumination of the PDC signal photon and the black-body radiation of 5777\,K. The molecular electronic states can only interact with the light at time $t \ge 0$. In the calculations, the average photon number shown in Fig.~\ref{fig:1} was employed, and the electronic transition energies were set to $\omega_{1{\rm g}} = 18000\,{\rm cm^{-1}} $ and $\omega_{2{\rm g}} = 18500\,{\rm cm^{-1}}$ with $\mu_{1g} = \mu_{2g}$. 
Because it is impossible to resolve the time at which the signal photon interacts with the molecule, it is considered that the signal photon interacts with the molecule at a uniformly random time. Consequently, the probability of observing the electronically excited molecule, $\rho_{11}(t) = \langle e_1 \vert \hat\rho_{\rm el}(t) \vert e_1 \rangle$, increases linearly with time. The same holds for the black-body radiation case, and therefore, the dynamics of the density matrix elements calculated for the two cases exhibit reasonably good agreement.
It should be noted that the probability does not continue to mount in the long-time limit when the radiative and nonradiative decays to the ground state are considered.
As demonstrated in Fig.~\ref{fig:2}, therefore, excited state dynamics induced by signal photons generated through PDC can be regarded as an emulation of the dynamics under the influence of sunlight irradiation, provided that $\bar{n}(\omega)$ reconstructs the spectrum of sunlight photons in a frequency range under investigation, such as the visible frequencies.
Figure~\ref{fig:2} also shows coherent time-evolution of the off-diagonal matrix element of the density operator, $\rho_{12}(t) = \langle e_1 \vert \hat\rho_{\rm el}(t) \vert e_2 \rangle$, although such coherent oscillations would be washed out by the excitation at uniformly random time \cite{Mancal:2010kc,Chenu:2014gr}. It is noted that the coherent oscillations can appear when the ``sudden turn-on of light'' is applied \cite{Chenu:2016cx,Tscherbul:2014cf}.

\begin{figure}
    \includegraphics{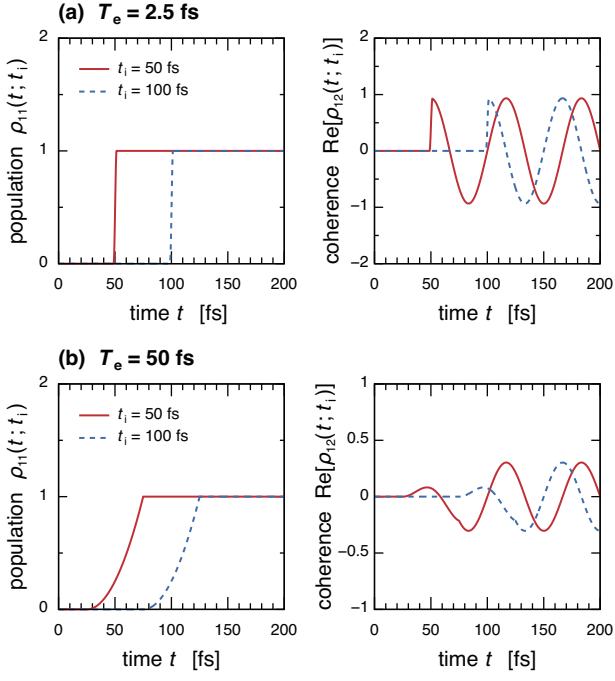}
    \caption{
	Time evolution of the normalized density matrix elements describing the electronic excitations triggered by interaction with the signal photons, provided that idler photons are detected at time $t=t_{\rm i}$,  Eq.~\eqref{eq:conditional-RDM}. 
	(a) The entanglement time $T_{\rm e}= 2.5\,{\rm fs}$ and the other parameters are the same as in Figs.~\ref{fig:1} and \ref{fig:2}. 
	(b) The entanglement time is set to $T_{\rm e} = 50\,{\rm fs}$, and the other parameters are $\bar{\omega}_{\rm s}=18001\,{\rm cm^{-1}}$ and $B=0.11$. 
	The normalization is such that the maximum value of the diagonal element, $\rho_{11}(t; t_{\rm i})=\langle e_1 \vert \hat\rho_{\rm el}(t; t_{\rm i}) \vert e_1 \rangle$, is unity.
    }
    \label{fig:3}
\end{figure}

\section{Detection of idler photons\label{sec:detection_of_idler}}

In the previous section, we discussed the interaction between the molecule and the signal photons without detecting the idler. However, more useful application of the PDC source can be made possible through detection of idler photons. Accordingly, spectroscopic and imaging techniques with entangled photon pairs have been proposed on the basis of coincidence counting \cite{Pittman:1995un,Yabushita:2004hy,Lee:2006id,Kalachev:2013kh,Raymer:2013kj,Dorfman:2014bn,Schlawin:2016er}. When the quantum states of both signal and idler photons are measured, characteristic features of quantum lights, such as entanglement time, can provide novel and useful control knobs to supplement classical parameters such as frequency and time delay. 
However, such heralded signal photons are not in the thermal state in contrast to the discussions in the preceding sections.
In the following, we investigate the excited state dynamics triggered by interaction of molecules with signal photons when idler photons are detected.

\begin{figure}
    \includegraphics{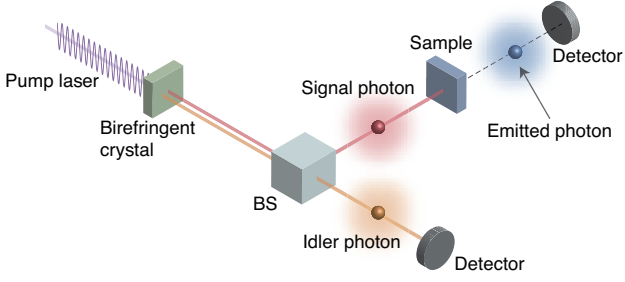}
    \caption{
    Illustration of the two photon coincidence counting measurement. The signal and idler beams generated through the PDC in the birefringent crystal are split on a beam splitter (BS). Only the signal photons interact with molecules, and the idler photons propagate freely. The idler photons and photons spontaneously emitted from the molecule excited by the signal photons are detected in coincidence. 
    }
    \label{fig:4}
\end{figure}

The optical length between the detector for idler photons and the PDC crystal is set to be the same as the length between the crystal and the sample into which signal photons enter. The photon detection that resolves the arrival time $t=t_{\rm i}$ of the idler photons is modeled with the projection operator: $\hat{\Pi}_{\rm i} (t_{\rm i}) = \hat{a}_{\rm i}^\dagger (t_{\rm i}) \lvert {\rm vac} \rangle \langle {\rm vac} \rvert \hat{a}_{\rm i} (t_{\rm i})$, where $\hat{a}_{\rm i}(t) = (2\pi)^{-1/2}\int d\omega\, \hat{a}_{\rm i}(\omega) e^{-i\omega t}$ has been introduced \cite{Du:2015gc}. The spectral information on the idler photon is not obtained. Consequently, the quantum state of the signal photon is a mixed state in terms of frequency, although the signal photon number is identified as unity. The frequency distribution is given by $D(\omega)\propto P_\omega(1) \simeq \tanh^2 r(\omega)$. In the CW pumping case, the first photon arrives at a uniformly random time \cite{Du:2015gc} and the second photon certainly arrives within the entanglement time $T_{\rm e}$.
Thus, the probability of detecting a photon in the idler beam is time-independent: $P_{\rm i} = \langle \psi_{\rm PDC} \rvert \hat{\Pi}_{\rm i}(t_{\rm i}) \lvert \psi_{\rm PDC} \rangle = Z^{-1} \int d\omega \tanh^2 r(\omega)$. When the idler photon is detected at time $t_{\rm i}$, the density operator of the electronic excitation is given by $\hat{\rho}_{\rm el} (t ; t_{\rm i}) = (1/P_{\rm i}) {\rm tr}_{{\rm s}+{\rm i}} [\hat{\Pi}_{\rm i}  (t_{\rm i}) \lvert \psi_{\rm total} (t) \rangle \langle \psi_{\rm total} (t) \rvert ]$, leading to an expression different from Eq.~\eqref{eq:unconditional-RDM1}:
\begin{align}
	\hat{\rho}_{\rm el}(t; t_{\rm i})
	&=
	\sum_{\alpha \beta}
	\frac{\mu_{\alpha g} \mu_{\beta g}}{\hbar^2}
	e^{-i\omega_{\alpha \beta} t}
	\int^t_{-\infty} d\tau_2\,
	e^{-i\omega_{\beta g} \tau_2}
\notag\\
    &\quad \times
	\int^t_{-\infty} d\tau_1\,
	e^{i\omega_{\alpha g} \tau_1}
	G_{\rm s}^{(1)} (\tau_2, \tau_1 ; t_{\rm i})
	\lvert e_\alpha \rangle\langle e_\beta \rvert,
	\label{eq:conditional-RDM}
\end{align}
where $ G_{\rm s}^{(1)}(t_2,t_1 ; t_{\rm i}) = \langle \hat{E}_{\rm s}^{(-)} (t_2) \hat{E}_{\rm s}^{(+)} (t_1) \rangle_{t_{\rm i}}$ is the first-order temporal correlation function of the heralded signal photon. The bracket represents $\langle \dots \rangle_{t_{\rm i}} = {\rm tr}_{\rm s+ i} [\dots \hat{\rho}_{\rm s}(t_{\rm i})]$, where $\hat{\rho}_{\rm s} (t_{\rm i})$ is the reduced density operator of the heralded signal photon given by $ \hat{\rho}_{\rm s} (t_{\rm i}) = P_{\rm i}^{-1} {\rm tr}_{\rm i}[ \hat{\Pi}_{\rm i} (t_{\rm i}) \lvert \psi_{\rm PDC} \rangle \langle \psi_{\rm PDC} \rvert]$. A concrete expression of the correlation function is obtained as
\begin{align}
    G_{\rm s}^{(1)}(t_2,t_1 ; t_{\rm i})
    =
    [\mathcal{E}^{(t_{\rm i})}(t_2)]^\ast
    \mathcal{E}^{(t_{\rm i})}(t_1),
    \label{eq:Glauber-idler-time}
\end{align}
where the following quantity has been introduced:
\begin{align}
	\mathcal{E}^{(t_{\rm i} )} (t)
	=
	\frac{1}{\sqrt{P_{\rm i} Z}}
	\int^\infty_0 d\omega\,
	e^{- i \omega(t -t_{\rm i})}
	A (\omega)
	\tanh r(\omega).
	\label{eq:electric-field}
\end{align}
Equation~\eqref{eq:electric-field} could be regarded as the ``electric field'' that will interact with the molecule.
Figure~\ref{fig:3} presents the time evolution of the density matrix elements under the condition that the idler photon is detected at time $t=t_{\rm i}$ for two values of the entanglement time, (a) $T_{\rm e} = 2.5\,{\rm fs}$ and (b) $T_{\rm e} = 50\,{\rm fs}$. The parameters chosen for Fig.~\ref{fig:3}a are the same as in Fig.~\ref{fig:2}, while the parameters employed in Fig.~\ref{fig:3}b are $\bar\omega_{\rm s}=18001\,{\rm cm^{-1}}$ and $B=0.11$. In contrast to the case in Fig.~\ref{fig:2}, the detection of the idler photon enables us to assign the time at which the signal photon interacts with the molecule. Thus, the probability of observing the electronically excited molecule, $\rho_{11}(t; t_{\rm i}) = \langle e_1 \vert \hat\rho_{\rm el} (t; t_{\rm i}) \vert e_1 \rangle$, exhibits the plateau values 0 and 1. However, the rise time from 0 to 1 depends on values of the entanglement time. This can be understood through the following approximative treatment of the ``electric field.''
From Eq.~\eqref{eq:photon-number-cw}, the PDC to generate light reproducing the weak intensity of sunlight lies in the weak down-conversion regime, $ r(\omega) \ll 1$. In this limit, the approximation of $\tanh r(\omega) \simeq r(\omega) = B\, {\rm sinc}[ (\omega - \bar{\omega}_{\rm s}) T_{\rm e}/2]$ is relevant 
and Eq.~\eqref{eq:electric-field} is recast as
\begin{align}
    \mathcal{E}^{(t_{\rm i})} (t)
    \propto
	A (\bar{\omega}_s)
	\frac{2\pi}{T_{\rm e}}
	{\rm rect} \left( \frac{t - t_{\rm i} }{T_{\rm e}} \right)
	e^{-i\bar{\omega}_{\rm s}(t-t_{\rm i})},
	\label{eq:temporal-correlation-CW}
\end{align}
where ${\rm rect}(x) = 1$ for $\lvert x \rvert <1/2$ and 0 otherwise.
While deriving Eq.~\eqref{eq:temporal-correlation-CW}, the approximation of $A(\omega)r(\omega) \simeq A(\bar\omega_{\rm s})r(\omega) $ is employed, where  $A(\omega)\propto \sqrt{\omega}$. This does not cause a fatal defect for the parameters employed in Fig.~\ref{fig:3}. 
Equation~\eqref{eq:temporal-correlation-CW} demonstrates that the signal photon certainly arrives at the molecular sample within the entanglement time $T_{\rm e}$ before or after the idler photon is detected at time $t_{\rm i}$. 
For $t > t_{\rm i} + T_{\rm e}/2$, the density operator in Eq.~\eqref{eq:conditional-RDM} is obtained as 
$
    \hat{\rho}_{\rm el}(t; t_{\rm i})
    \propto 
    \sum_{\alpha \beta} \mu_{\alpha g}\mu_{\beta g}
    {\rm sinc} [(\omega_{\alpha g} - \bar{\omega}_{\rm s}) T_{\rm e}/2]
    {\rm sinc} [(\omega_{\beta g} - \bar{\omega}_{\rm s}) T_{\rm e}/2]
      e^{-i\omega_{\alpha\beta}(t-t_{\rm i})}
      \lvert e_\alpha \rangle\langle e_\beta \rvert
$.
When the entanglement time $T_{\rm e}$ is extremely short, Eq.~\eqref{eq:temporal-correlation-CW} can be approximated as $ \mathcal{E}^{(t_{\rm i})}(t) \propto \delta(t-t_{\rm i})$. In this ``impulsive'' limit, the time evolution of the density operator of the electronic excitation, Eq.~\eqref{eq:conditional-RDM} reduces to $ \hat{\rho}_{\rm el}(t; t_{\rm i}) \propto \sum_{\alpha \beta} \mu_{\alpha g}\mu_{\beta g} e^{-i\omega_{\alpha\beta} (t-t_{\rm i})} \lvert e_\alpha \rangle\langle e_\beta \rvert $, as presented in Fig.~\ref{fig:3}a. 
Therefore, it is reasonable to state that the detection time for the idler photon is considered to be the time that the signal photon arrives at the molecule in the case of short entanglement time.
In contrast, the arrival time of the signal photon becomes blurred within the time window of $t_{\rm i} - T_{\rm e}/2 \le t \le t_{\rm i} + T_{\rm e}/2$, when the entanglement time $T_{\rm e}$ is relatively long.
This situation is presented in Fig.~\ref{fig:3}b.
Experimentally the dynamics described with Eq.~\eqref{eq:conditional-RDM} could be observed through two-photon coincidence counting \cite{Dorfman:2016da,Glauber2007book} although the actual measurement might be technically difficult.
As illustrated in Fig.~\ref{fig:4}, idler photons and photons spontaneously emitted from the molecule excited by signal photons are counted at times $t_{\rm i}$ and $t$, respectively. 
The two-photon counting signal is thus given by $S(t,t_{\rm i}) = {\rm tr}[ \hat{a}_{\rm i}^\dagger(t_{\rm i})\hat{a}_{\rm s}^\dagger(t)\hat{a}_{\rm s}(t)\hat{a}_{\rm i}(t_{\rm i}) \hat\rho_{\rm tot}(t)]$, which is approximately expressed with the density matrices in Eq.~\eqref{eq:conditional-RDM} as
\begin{align}
	S(t,t_i)
	\simeq
	\frac{2\pi}{\hbar^{2}}
	A(\bar\omega_{\rm s})^2
	\sum_{\alpha\beta}
	\mu_{\alpha g}\mu_{\beta g}
	\langle e_\alpha \vert 
		\hat\rho_{\rm el}(t;t_i)
	\vert e_\beta \rangle.
	\label{eq:TPC-signal}
\end{align}
This expression is derived in Appendix~A.

As aforementioned, idler photons are detected at a uniformly random time in the CW pumping case.
To gain further insight into physical implications of Eq.~\eqref{eq:conditional-RDM} and correspondingly Eq.~\eqref{eq:TPC-signal}, we consider the average of the density operator
of the electronic excitation, Eq.~\eqref{eq:conditional-RDM}, for all possible values of
$t_{\rm i}$,
\begin{align}
    \langle \hat{\rho}_{\rm el}(t;t_{\rm i}) \rangle
    &=
	\sum_{\alpha \beta}
	\frac{\mu_{\alpha g} \mu_{\beta g}}{\hbar^2}
	e^{-i\omega_{\alpha \beta} t}
	\int^t_{-\infty} d\tau_2\,
	e^{-i\omega_{\beta g} \tau_2}
\notag\\
    &\quad \times
	\int^t_{-\infty} d\tau_1\,
	e^{i\omega_{\alpha g} \tau_1}
	\langle
	    G_{\rm s}^{(1)}(\tau_2,\tau_1 ; t_{\rm i})
    \rangle
    \lvert e_\alpha \rangle\langle e_\beta \rvert
    \label{eq:averaged-RDM}
\end{align}
with 
$\langle G_{\rm s}^{(1)}(t_2,t_1 ; t_{\rm i}) \rangle$ being the correlation function
averaged in terms of $t_{\rm i}$, 
\begin{align}
	\langle G_{\rm s}^{(1)}(t_2,t_1 ; t_{\rm i}) \rangle
	 \propto \int^\infty_{-\infty} dt_{\rm i}\, [\mathcal{E}^{(t_{\rm i})}(t_2)]^\ast \mathcal{E}^{(t_{\rm i})}(t_1).
\end{align}
As illustrated in Fig.~\ref{fig:1}, the photon number in the visible region is substantially 0 or 1, and the average photon number $\bar{n}(\omega)$ is well approximated by
\begin{align}
    \bar{n}(\omega)
    \simeq 
    \sum_{n=0}^1 n P_\omega(n) 
    =
    P_{\omega}(1)
    \simeq 
    \tanh^2 r(\omega).
    \label{eq:photon-number-cw-approx}
\end{align}
As a consequence, Eq.~\eqref{eq:electric-field} is approximately expressed as
\begin{align}
	\mathcal{E}^{(t_{\rm i} )} (t)
	\simeq
    \frac{1}{\sqrt{P_{\rm i}Z}}
	\int^\infty_0 d\omega\,
	e^{- i \omega(t -t_{\rm i})}
	A (\omega)
	\sqrt{\bar{n}(\omega)},
	\label{eq:approximated-pseudo-field}
\end{align}
and the averaged correlation function is computed as 
\begin{align}
	\langle G_{\rm s}^{(1)}(t_2,t_1 ; t_{\rm i}) \rangle \propto  G_{\rm s}^{(1)}(t_2,t_1),
\end{align}
indicating that Eq.~\eqref{eq:averaged-RDM} is identical to Eq.~\eqref{eq:unconditional-RDM1}.
Indeed, the sample average of the density matrix elements presented in Fig.~\ref{fig:3} in terms of  $t_{\rm i}$ reproduces the curves depicted in Fig.~\ref{fig:2}
\footnote{The correlation function in Eq.~\eqref{eq:averaged-RDM} is practically identical to the first-order temporal correlation function for individual realizations of the transform-limited classical laser pulses investigated by Chenu and Brumer to yield the thermal light result \cite{Chenu:2016cx} when $\bar{n}(\omega)$ in Eq.~\eqref{eq:approximated-pseudo-field} is replaced with $\bar{n}_{\rm th}(\omega)$.}. 
What should be emphasized here is that the sample average in terms of $t_{\rm i}$ does not change physical properties of the heralded signal photon and the interaction with the molecule: it characterizes the statistical property of the heralded signal photon.
These observations indicate that the detection of the idler photon enables us to resolve the photon arrival times under the pseudo-sunlight irradiation. 
However, it should be also noted that this conclusion is only true for cases where Eq.~\eqref{eq:photon-number-cw-approx} is an acceptable approximation, e.g., in the solar visible region. In such cases, the photon number is substantially 0 or 1, and the heralding is to remove the vacuum state of the photon number 0 \cite{Jin:2011dt}\footnote{After submitting this paper, we became aware of Ref.~\cite{Deng:2019cu}, in which C.-Y. Lu, J.-W. Pan and coworkers experimentally demonstrated nonclassical interference between sunlight and single photons from a quantum dot. For measuring the Hong-Ou-Mandel interference, only the single-photon events of the sunlight were registered.}, which does not interact with the molecule. Therefore, the heralding is of no major consequence when considering the molecule-photon interaction.

\section{Concluding remarks}

In this work we theoretically demonstrated that the nature of sunlight photons can be emulated through quantum entangled photons generated with the PDC. One may emulate the sunlight, which is the radiation from the black-body with an effective temperature of approximately 5777\,K, through controlling the system's parameters in a mechanical fashion.
Further, electronic excitations of a molecule using such pseudo-sunlight light were investigated. The key is that the entanglement time, which is a unique characteristic of the quantum entangled photons, serves as a control knob to resolve the photon arrival times, enabling investigations on real-time dynamics triggered by the pseudo-sunlight photons. Pinpointing the photon arrival times may pave a new path for implementing time-resolved spectroscopic experiments that directly reflects properties of natural sunlight.

\begin{acknowledgements}
AI thanks Dr.~Yutaka Shikano for his informative comments. 
This work was supported by MEXT Quantum Leap Flagship Program Grant Number~JPMXS0118069242, JSPS KAKENHI Grant Numbers~17H02946 and 18H01937, and MEXT KAKENHI Grant Number~17H06437 in Innovative Areas ``Innovations for Light-Energy Conversion ($\rm I^4LEC$).''
\end{acknowledgements}

\appendix 
\section{Two photon coincidence counting signal\label{appendix}}

\renewcommand{\theequation}{\ref{appendix}\arabic{equation}}
\renewcommand{\thefigure}{\ref{appendix}\arabic{figure}}
\setcounter{equation}{0}
\setcounter{figure}{0}

We consider the two-photon coincidence counting measurement depicted in Fig.~\ref{fig:4}, in which the idler photons and the photons spontaneously emitted from the molecule excited by the signal photons are counted at times $t_{\rm i}$ and $t$, respectively. The two-photon counting signal is thus written as \cite{Glauber2007book,Dorfman:2016da} 
\begin{align}
	S(t,t_{\rm i})
	=
	\mathrm{tr} 
	\left[
		\hat{a}_{\rm i}^\dagger(t_{\rm i}) a_{\rm s}^\dagger(t) \hat{a}_{\rm s}(t) \hat{a}_{\rm i}(t_{\rm i}) \hat\rho_{\rm tot}(t)
	\right],
	\label{eq:two-photon-counting:0}
\end{align}
where the density operator $\hat\rho_{\rm tot}(t)$ represents the state of the total system after the spontaneous emission.
In this Appendix, we evaluate Eq.~\eqref{eq:two-photon-counting:0} to relate the two-photon counting signal with the electronically excited state dynamics induced by the signal photons under the condition that the idler photon is detected at time $t_{\rm i}$.

The density operator $\hat\rho_{\rm tot}(t)$ can be expanded up to forth-order with respect to $\hat{H}_{\rm int}(t)$ as \cite{Mukamel:1995us,Dorfman:2016da}
\begin{align}
	\hat\rho_{\rm tot}(t)
	&=
	\left(-\frac{i}{\hbar}\right)^4
	\int^t_{-\infty}d\tau_4 
	\int^{\tau_4}_{-\infty} d\tau_3
	\int^{\tau_3}_{-\infty} d\tau_2
	\int^{\tau_2}_{-\infty} d\tau_1
\notag \\
&\quad\times
	\hat{\mathcal{G}}(t-\tau_4) 
	\hat{H}_{\rm int}^\times(\tau_4)
	\hat{\mathcal{G}}(\tau_4-\tau_3) 
	\hat{H}_{\rm int}^\times(\tau_3)
	\hat{\mathcal{G}}(\tau_3-\tau_2) 
\notag \\
&\quad\times
	\hat{H}_{\rm int}^\times(\tau_2)
	\hat{\mathcal{G}}(\tau_2-\tau_1) 
	\hat{H}_{\rm int}^\times(\tau_1)
	\hat\rho_{\rm tot}(-\infty),
	\label{eq:4th-density-op:0}
\end{align}
where $\hat{\mathcal{G}}(t)$ denotes the  the Liouville space time-evolution operator to describe dynamics of the electronic excitation in the molecules, and the superoperator notation $\hat{O}_1^\times \hat{O}_2 = [\hat{O}_1, \hat{O}_2]$ has been introduced for any operators $\hat{O}_1$ and $\hat{O}_2$. The initial state is assumed to be $\hat\rho_{\rm tot}(-\infty) = \lvert g \rangle\langle g \rvert \otimes \lvert \psi_{\rm PDC} \rangle\langle \psi_{\rm PDC} \rvert$.
For simplicity, we consider the situation that there are no dissipative processes such as environmental effects and exciton relaxation. In this situation, the time-evolution operator can be written as $\hat{\mathcal{G}}(t)\hat{O} = \hat{G}(t)\hat{O}\hat{G}^\dagger(t)$ with $\hat{G}(t)= \exp (-i \hat{H}_{\rm mol}t / \hbar )$. Therefore, the density operator effective to the calculation of Eq.~\eqref{eq:two-photon-counting:0} is obtained as
\begin{align}
	\hat\rho_{\rm tot}(t)
	&=
	\left(-\frac{i}{\hbar}\right)^4
	\int^t_{-\infty}d\tau_2
	\int^{\tau_2}_{-\infty} d\tau_1
	\int^t_{-\infty} d\tau_2'
	\int^{\tau_2'}_{-\infty} d\tau_1'
\notag \\
&\quad\times
	\hat{G}(t-\tau_2) 
	\hat{H}_{\rm int}(\tau_2)
	\hat{G}(\tau_2-\tau_1) 
	\hat{H}_{\rm int}(\tau_1)
	\hat\rho_{\rm tot}(-\infty)
\notag \\
&\quad\times
	\hat{H}_{\rm int}(\tau_1')
	\hat{G}^\dagger(\tau_2'-\tau_1')
	\hat{H}_{\rm int}(\tau_2')
	\hat{G}^\dagger(t-\tau_2'). 
	\label{eq:4th-density-op:1}
\end{align}
Further, the rotating-wave approximation, $\hat{H}_{\rm int}(t) \simeq -\sum_\alpha\mu_{\alpha g} \lvert e_{\alpha} \rangle\langle g \rvert \hat{E}_{\rm s}^{(+)}(t) + {\rm h.c.}$, leads to
\begin{align}
	\hat\rho_{\rm tot}(t)
	&\simeq 
	\sum_{\alpha \beta}
	\frac{\mu_{\alpha g}^2 \mu_{\beta g}^2}{\hbar^4}
	\int^t_{-\infty}d\tau_2
	\int^{\tau_2}_{-\infty} d\tau_1
	\int^t_{-\infty} d\tau_2'
	\int^{\tau_2'}_{-\infty} d\tau_1'
\notag \\
	&\quad\times
	\hat{E}_{\rm s}^{(-)}(\tau_2)
	e^{-i\omega_{\alpha g}(\tau_2-\tau_1)}
	\hat{E}_{\rm s}^{(+)}(\tau_1)
	\vert e_\alpha \rangle
	\lvert \psi_{\rm PDC} \rangle
\notag \\
	&\quad\times
	\langle \psi_{\rm PDC} \rvert
	\langle e_\beta\vert
	\hat{E}_{\rm s}^{(-)}(\tau_1')
	e^{i\omega_{\beta g}(\tau_2'-\tau_1')}
	\hat{E}_{\rm s}^{(+)}(\tau_2').
	\label{eq:4th-density-op:2}
\end{align}
Hence, the two-photon coincidence counting signal in Eq.~\eqref{eq:two-photon-counting:0} is obtained as
\begin{align}
	S(t,t_{\rm i})
	&=
	\sum_{\alpha \beta}
	\frac{\mu_{\alpha g}^2\mu_{\beta g}^2}{\hbar^4}
	\int^t_{-\infty}d\tau_2  \int^{\tau_2}_{-\infty} d\tau_1
	\int^t_{-\infty}d\tau_2' \int^{\tau_2'}_{-\infty} d\tau_1'
\notag \\
&\quad\times
	e^{i\omega_{\beta g}(\tau_2'-\tau_1')}
	e^{-i\omega_{\alpha g}(\tau_2-\tau_1)}
	D(t,t_{\rm i}; \tau_2,\tau_1,\tau_2'\tau_1'),
\end{align}
where $D(t,t_{\rm i}; \tau_2,\tau_1,\tau_2'\tau_1')$ is the multi-point correlation function of the electric field operators and the creation/annihilation operators of the photons,
\begin{align}
	D(t,t_{\rm i}; \tau_2,\tau_1,\tau_2',\tau_1')
	&=
	\langle \psi_{\rm PDC} \vert 
		\hat{E}_{\rm s}^{(-)}(\tau_1')
		\hat{E}_{\rm s}^{(+)}(\tau_2') 
		\hat{a}_{\rm i}^\dagger(t_i) \hat{a}_{\rm s}^\dagger(t)
		\hat{a}_{\rm s}(t)
\notag\\
    &\quad \times
		\hat{a}_{\rm i}(t_{\rm i}) 
		\hat{E}_{\rm s}^{(-)}(\tau_2)
		\hat{E}_{\rm s}^{(+)}(\tau_1)
	\vert \psi_{\rm PDC} \rangle. 
	\label{eq:multipoint-correlation}
\end{align}
When the electric field operators are approximately expressed as $\hat{E}_{\rm s}^{(+)}(t) \simeq i A(\bar\omega_{\rm s})\int d\omega\,\hat{a}_{\rm s}(\omega)e^{-i\omega t}$ and $\hat{E}_{\rm s}^{(-)}(t) \simeq -i A(\bar\omega_{\rm s})\int d\omega\,\hat{a}_{\rm s}^\dagger(\omega)e^{i\omega t}$, 
 the following commutation relations are obtained:
\begin{align}
	[\hat{a}_{\rm s}^\dagger(t) \hat{a}_{\rm s}(t) , \hat{E}_{\rm s}^{(-)}(\tau)]
	&\simeq
	\hat{E}_{\rm s}^{(-)}(t)\delta(t-\tau),
\\
	[\hat{E}_{\rm s}^{(+)}(t), \hat{E}_{\rm s}^{(-)}(\tau)]
	&\simeq
	2\pi  A(\bar\omega_{\rm s})^2 \delta(t-\tau),
\end{align}
which enable us to calculate Eq.~\eqref{eq:multipoint-correlation} as
\begin{align}
	D(t,t_{\rm i}; \tau_2,\tau_1,\tau_2',\tau_1')
	\simeq
	2\pi A(\bar\omega_{\rm s})^2 \delta(t-\tau_2) \delta(t-\tau_2')
	G_{\rm s}^{(1)}(\tau_1',\tau_1;t_{\rm i}).
\end{align}
As the consequence, we obtain the expression of the two-photon coincidence counting signal as
\begin{align}
	S(t,t_{\rm i})
	&\simeq
	\frac{2\pi}{\hbar^4}
	A(\bar\omega_{\rm s})^2
	\sum_{\alpha \beta}\mu_{\alpha g}^2\mu_{\beta g}^2
	\int^{t}_{-\infty} d\tau_1
	\int^{t}_{-\infty} d\tau_1'
\notag\\
    &\quad \times
	e^{i\omega_{\beta g}(t-\tau_1')}
	e^{-i\omega_{\alpha g}(t-\tau_1)}
	G_{\rm s}^{(1)}(\tau_1',\tau_1;t_{\rm i}),
	\label{eq:two-photon-counting}
\end{align}
which is recast into a simpler form with the use of $\hat\rho_{\rm el}(t;t_{\rm i})$,
\begin{align}
	S(t,t_{\rm i})
	\simeq
	\frac{2\pi }{\hbar^2}A(\bar\omega_{\rm s})^2
	\sum_{\alpha \beta}
	\mu_{\alpha g}\mu_{\beta g}
	\langle e_\alpha\vert \hat\rho_{\rm el}(t;t_{\rm i})\vert e_\beta \rangle.
	\label{eq:two-photon-counting-final}
\end{align}
Equation~\eqref{eq:two-photon-counting-final} indicates that the electronically excited state dynamics in Eq.~(6) can be observed through the two-photon coincidence measurement.


\begin{thebibliography}{44}%
\makeatletter
\providecommand \@ifxundefined [1]{%
 \@ifx{#1\undefined}
}%
\providecommand \@ifnum [1]{%
 \ifnum #1\expandafter \@firstoftwo
 \else \expandafter \@secondoftwo
 \fi
}%
\providecommand \@ifx [1]{%
 \ifx #1\expandafter \@firstoftwo
 \else \expandafter \@secondoftwo
 \fi
}%
\providecommand \natexlab [1]{#1}%
\providecommand \enquote  [1]{``#1''}%
\providecommand \bibnamefont  [1]{#1}%
\providecommand \bibfnamefont [1]{#1}%
\providecommand \citenamefont [1]{#1}%
\providecommand \href@noop [0]{\@secondoftwo}%
\providecommand \href [0]{\begingroup \@sanitize@url \@href}%
\providecommand \@href[1]{\@@startlink{#1}\@@href}%
\providecommand \@@href[1]{\endgroup#1\@@endlink}%
\providecommand \@sanitize@url [0]{\catcode `\\12\catcode `\$12\catcode
  `\&12\catcode `\#12\catcode `\^12\catcode `\_12\catcode `\%12\relax}%
\providecommand \@@startlink[1]{}%
\providecommand \@@endlink[0]{}%
\providecommand \url  [0]{\begingroup\@sanitize@url \@url }%
\providecommand \@url [1]{\endgroup\@href {#1}{\urlprefix }}%
\providecommand \urlprefix  [0]{URL }%
\providecommand \Eprint [0]{\href }%
\providecommand \doibase [0]{http://dx.doi.org/}%
\providecommand \selectlanguage [0]{\@gobble}%
\providecommand \bibinfo  [0]{\@secondoftwo}%
\providecommand \bibfield  [0]{\@secondoftwo}%
\providecommand \translation [1]{[#1]}%
\providecommand \BibitemOpen [0]{}%
\providecommand \bibitemStop [0]{}%
\providecommand \bibitemNoStop [0]{.\EOS\space}%
\providecommand \EOS [0]{\spacefactor3000\relax}%
\providecommand \BibitemShut  [1]{\csname bibitem#1\endcsname}%
\let\auto@bib@innerbib\@empty
\bibitem [{\citenamefont {Cheng}\ and\ \citenamefont
  {Fleming}(2009)}]{Cheng:2009co}%
  \BibitemOpen
  \bibfield  {author} {\bibinfo {author} {\bibfnamefont {Y.-C.}\ \bibnamefont
  {Cheng}}\ and\ \bibinfo {author} {\bibfnamefont {G.~R.}\ \bibnamefont
  {Fleming}},\ }\href@noop {} {\bibfield  {journal} {\bibinfo  {journal} {Annu.
  Rev. Phys. Chem.}\ }\textbf {\bibinfo {volume} {60}},\ \bibinfo {pages} {241}
  (\bibinfo {year} {2009})}\BibitemShut {NoStop}%
\bibitem [{\citenamefont {Man{\v c}al}\ and\ \citenamefont
  {Valkunas}(2010)}]{Mancal:2010kc}%
  \BibitemOpen
  \bibfield  {author} {\bibinfo {author} {\bibfnamefont {T.}~\bibnamefont
  {Man{\v c}al}}\ and\ \bibinfo {author} {\bibfnamefont {L.}~\bibnamefont
  {Valkunas}},\ }\href@noop {} {\bibfield  {journal} {\bibinfo  {journal} {New
  J. Phys.}\ }\textbf {\bibinfo {volume} {12}},\ \bibinfo {pages} {065044}
  (\bibinfo {year} {2010})}\BibitemShut {NoStop}%
\bibitem [{\citenamefont {Ishizaki}\ and\ \citenamefont
  {Fleming}(2011)}]{Ishizaki:2011cx}%
  \BibitemOpen
  \bibfield  {author} {\bibinfo {author} {\bibfnamefont {A.}~\bibnamefont
  {Ishizaki}}\ and\ \bibinfo {author} {\bibfnamefont {G.~R.}\ \bibnamefont
  {Fleming}},\ }\href@noop {} {\bibfield  {journal} {\bibinfo  {journal} {J.
  Phys. Chem. B}\ }\textbf {\bibinfo {volume} {115}},\ \bibinfo {pages} {6227}
  (\bibinfo {year} {2011})}\BibitemShut {NoStop}%
\bibitem [{\citenamefont {Brumer}\ and\ \citenamefont
  {Shapiro}(2012)}]{Brumer:2012ib}%
  \BibitemOpen
  \bibfield  {author} {\bibinfo {author} {\bibfnamefont {P.}~\bibnamefont
  {Brumer}}\ and\ \bibinfo {author} {\bibfnamefont {M.}~\bibnamefont
  {Shapiro}},\ }\href@noop {} {\bibfield  {journal} {\bibinfo  {journal} {Proc.
  Natl. Acad. Sci. USA}\ }\textbf {\bibinfo {volume} {109}},\ \bibinfo {pages}
  {19575} (\bibinfo {year} {2012})}\BibitemShut {NoStop}%
\bibitem [{\citenamefont {Fassioli}\ \emph {et~al.}(2012)\citenamefont
  {Fassioli}, \citenamefont {Olaya-Castro},\ and\ \citenamefont
  {Scholes}}]{Fassioli:2012gd}%
  \BibitemOpen
  \bibfield  {author} {\bibinfo {author} {\bibfnamefont {F.}~\bibnamefont
  {Fassioli}}, \bibinfo {author} {\bibfnamefont {A.}~\bibnamefont
  {Olaya-Castro}}, \ and\ \bibinfo {author} {\bibfnamefont {G.~D.}\
  \bibnamefont {Scholes}},\ }\href@noop {} {\bibfield  {journal} {\bibinfo
  {journal} {J. Phys. Chem. Lett.}\ }\textbf {\bibinfo {volume} {3}},\ \bibinfo
  {pages} {3136} (\bibinfo {year} {2012})}\BibitemShut {NoStop}%
\bibitem [{\citenamefont {Chenu}\ \emph
  {et~al.}(2015{\natexlab{a}})\citenamefont {Chenu}, \citenamefont
  {Bra{\'n}czyk}, \citenamefont {Scholes},\ and\ \citenamefont
  {Sipe}}]{Chenu:2015hi}%
  \BibitemOpen
  \bibfield  {author} {\bibinfo {author} {\bibfnamefont {A.}~\bibnamefont
  {Chenu}}, \bibinfo {author} {\bibfnamefont {A.~M.}\ \bibnamefont
  {Bra{\'n}czyk}}, \bibinfo {author} {\bibfnamefont {G.~D.}\ \bibnamefont
  {Scholes}}, \ and\ \bibinfo {author} {\bibfnamefont {J.~E.}\ \bibnamefont
  {Sipe}},\ }\href@noop {} {\bibfield  {journal} {\bibinfo  {journal} {Phys.
  Rev. Lett.}\ }\textbf {\bibinfo {volume} {114}},\ \bibinfo {pages} {213601}
  (\bibinfo {year} {2015}{\natexlab{a}})}\BibitemShut {NoStop}%
\bibitem [{\citenamefont {Brumer}(2018)}]{Brumer:2018ka}%
  \BibitemOpen
  \bibfield  {author} {\bibinfo {author} {\bibfnamefont {P.}~\bibnamefont
  {Brumer}},\ }\href@noop {} {\bibfield  {journal} {\bibinfo  {journal} {J.
  Phys. Chem. Lett.}\ }\textbf {\bibinfo {volume} {9}},\ \bibinfo {pages}
  {2946} (\bibinfo {year} {2018})}\BibitemShut {NoStop}%
\bibitem [{\citenamefont {Chan}\ \emph {et~al.}(2018)\citenamefont {Chan},
  \citenamefont {Gamel}, \citenamefont {Fleming},\ and\ \citenamefont
  {Whaley}}]{Chan:2018em}%
  \BibitemOpen
  \bibfield  {author} {\bibinfo {author} {\bibfnamefont {H.~C.}\ \bibnamefont
  {Chan}}, \bibinfo {author} {\bibfnamefont {O.~E.}\ \bibnamefont {Gamel}},
  \bibinfo {author} {\bibfnamefont {G.~R.}\ \bibnamefont {Fleming}}, \ and\
  \bibinfo {author} {\bibfnamefont {K.~B.}\ \bibnamefont {Whaley}},\
  }\href@noop {} {\bibfield  {journal} {\bibinfo  {journal} {J. Phys. B: At.
  Mol. Opt. Phys.}\ }\textbf {\bibinfo {volume} {51}},\ \bibinfo {pages}
  {054002} (\bibinfo {year} {2018})}\BibitemShut {NoStop}%
\bibitem [{\citenamefont {Shatokhin}\ \emph {et~al.}(2018)\citenamefont
  {Shatokhin}, \citenamefont {Walschaers}, \citenamefont {Schlawin},\ and\
  \citenamefont {Buchleitner}}]{Shatokhin:2018jq}%
  \BibitemOpen
  \bibfield  {author} {\bibinfo {author} {\bibfnamefont {V.~N.}\ \bibnamefont
  {Shatokhin}}, \bibinfo {author} {\bibfnamefont {M.}~\bibnamefont
  {Walschaers}}, \bibinfo {author} {\bibfnamefont {F.}~\bibnamefont
  {Schlawin}}, \ and\ \bibinfo {author} {\bibfnamefont {A.}~\bibnamefont
  {Buchleitner}},\ }\href@noop {} {\bibfield  {journal} {\bibinfo  {journal}
  {New J. Phys.}\ }\textbf {\bibinfo {volume} {20}},\ \bibinfo {pages} {113040}
  (\bibinfo {year} {2018})}\BibitemShut {NoStop}%
\bibitem [{\citenamefont {Mu{\~n}oz}\ and\ \citenamefont
  {Schlawin}()}]{muoz2019photon}%
  \BibitemOpen
  \bibfield  {author} {\bibinfo {author} {\bibfnamefont {C.~S.}\ \bibnamefont
  {Mu{\~n}oz}}\ and\ \bibinfo {author} {\bibfnamefont {F.}~\bibnamefont
  {Schlawin}},\ }\href@noop {} {}\Eprint {http://arxiv.org/abs/1911.05054}
  {arXiv:1911.05054 [quant-ph]} \BibitemShut {NoStop}%
\bibitem [{\citenamefont {Mandel}\ and\ \citenamefont
  {Wolf}(1995)}]{mandel1995optical}%
  \BibitemOpen
  \bibfield  {author} {\bibinfo {author} {\bibfnamefont {L.}~\bibnamefont
  {Mandel}}\ and\ \bibinfo {author} {\bibfnamefont {E.}~\bibnamefont {Wolf}},\
  }\href@noop {} {\emph {\bibinfo {title} {Optical Coherence and Quantum
  Optics}}}\ (\bibinfo  {publisher} {Cambridge University Press},\ \bibinfo
  {address} {Cambridge},\ \bibinfo {year} {1995})\BibitemShut {NoStop}%
\bibitem [{\citenamefont {Estes}\ \emph {et~al.}(1971)\citenamefont {Estes},
  \citenamefont {Narducci},\ and\ \citenamefont {Tuft}}]{Estes:1971if}%
  \BibitemOpen
  \bibfield  {author} {\bibinfo {author} {\bibfnamefont {L.~E.}\ \bibnamefont
  {Estes}}, \bibinfo {author} {\bibfnamefont {L.~M.}\ \bibnamefont {Narducci}},
  \ and\ \bibinfo {author} {\bibfnamefont {R.~A.}\ \bibnamefont {Tuft}},\
  }\href@noop {} {\bibfield  {journal} {\bibinfo  {journal} {J. Opt. Soc. Am.}\
  }\textbf {\bibinfo {volume} {61}},\ \bibinfo {pages} {1301} (\bibinfo {year}
  {1971})}\BibitemShut {NoStop}%
\bibitem [{\citenamefont {Turner}\ \emph {et~al.}(2013)\citenamefont {Turner},
  \citenamefont {Arpin}, \citenamefont {McClure}, \citenamefont {Ulness},\ and\
  \citenamefont {Scholes}}]{Turner:2013fq}%
  \BibitemOpen
  \bibfield  {author} {\bibinfo {author} {\bibfnamefont {D.~B.}\ \bibnamefont
  {Turner}}, \bibinfo {author} {\bibfnamefont {P.~C.}\ \bibnamefont {Arpin}},
  \bibinfo {author} {\bibfnamefont {S.~D.}\ \bibnamefont {McClure}}, \bibinfo
  {author} {\bibfnamefont {D.~J.}\ \bibnamefont {Ulness}}, \ and\ \bibinfo
  {author} {\bibfnamefont {G.~D.}\ \bibnamefont {Scholes}},\ }\href@noop {}
  {\bibfield  {journal} {\bibinfo  {journal} {Nat. Commun.}\ }\textbf {\bibinfo
  {volume} {4}},\ \bibinfo {pages} {2298} (\bibinfo {year} {2013})}\BibitemShut
  {NoStop}%
\bibitem [{\citenamefont {Chenu}\ \emph
  {et~al.}(2015{\natexlab{b}})\citenamefont {Chenu}, \citenamefont
  {Bra{\'n}czyk},\ and\ \citenamefont {Sipe}}]{Chenu:2015km}%
  \BibitemOpen
  \bibfield  {author} {\bibinfo {author} {\bibfnamefont {A.}~\bibnamefont
  {Chenu}}, \bibinfo {author} {\bibfnamefont {A.~M.}\ \bibnamefont
  {Bra{\'n}czyk}}, \ and\ \bibinfo {author} {\bibfnamefont {J.~E.}\
  \bibnamefont {Sipe}},\ }\href@noop {} {\bibfield  {journal} {\bibinfo
  {journal} {Phys. Rev. A}\ }\textbf {\bibinfo {volume} {91}},\ \bibinfo
  {pages} {063813} (\bibinfo {year} {2015}{\natexlab{b}})}\BibitemShut
  {NoStop}%
\bibitem [{\citenamefont {Chenu}\ and\ \citenamefont
  {Brumer}(2016)}]{Chenu:2016cx}%
  \BibitemOpen
  \bibfield  {author} {\bibinfo {author} {\bibfnamefont {A.}~\bibnamefont
  {Chenu}}\ and\ \bibinfo {author} {\bibfnamefont {P.}~\bibnamefont {Brumer}},\
  }\href@noop {} {\bibfield  {journal} {\bibinfo  {journal} {J. Chem. Phys.}\
  }\textbf {\bibinfo {volume} {144}},\ \bibinfo {pages} {044103} (\bibinfo
  {year} {2016})}\BibitemShut {NoStop}%
\bibitem [{\citenamefont {Barnett}\ and\ \citenamefont
  {Knight}(1985)}]{Barnett:1985jc}%
  \BibitemOpen
  \bibfield  {author} {\bibinfo {author} {\bibfnamefont {S.~M.}\ \bibnamefont
  {Barnett}}\ and\ \bibinfo {author} {\bibfnamefont {P.~L.}\ \bibnamefont
  {Knight}},\ }\href@noop {} {\bibfield  {journal} {\bibinfo  {journal} {J.
  Opt. Soc. Am. B}\ }\textbf {\bibinfo {volume} {2}},\ \bibinfo {pages} {467}
  (\bibinfo {year} {1985})}\BibitemShut {NoStop}%
\bibitem [{\citenamefont {Yurke}\ and\ \citenamefont
  {Potasek}(1987)}]{Yurke:1987hs}%
  \BibitemOpen
  \bibfield  {author} {\bibinfo {author} {\bibfnamefont {B.}~\bibnamefont
  {Yurke}}\ and\ \bibinfo {author} {\bibfnamefont {M.}~\bibnamefont
  {Potasek}},\ }\href@noop {} {\bibfield  {journal} {\bibinfo  {journal} {Phys.
  Rev. A}\ }\textbf {\bibinfo {volume} {36}},\ \bibinfo {pages} {3464}
  (\bibinfo {year} {1987})}\BibitemShut {NoStop}%
\bibitem [{\citenamefont {Gerry}\ and\ \citenamefont
  {Knight}(2005)}]{Gerry2005:introductory}%
  \BibitemOpen
  \bibfield  {author} {\bibinfo {author} {\bibfnamefont {C.}~\bibnamefont
  {Gerry}}\ and\ \bibinfo {author} {\bibfnamefont {P.}~\bibnamefont {Knight}},\
  }\href@noop {} {\emph {\bibinfo {title} {Introductory Quantum Optics}}}\
  (\bibinfo  {publisher} {Cambridge University Press},\ \bibinfo {address}
  {Cambridge},\ \bibinfo {year} {2005})\BibitemShut {NoStop}%
\bibitem [{\citenamefont {Christ}\ \emph {et~al.}(2013)\citenamefont {Christ},
  \citenamefont {Brecht}, \citenamefont {Mauerer},\ and\ \citenamefont
  {Silberhorn}}]{Christ:2013fg}%
  \BibitemOpen
  \bibfield  {author} {\bibinfo {author} {\bibfnamefont {A.}~\bibnamefont
  {Christ}}, \bibinfo {author} {\bibfnamefont {B.}~\bibnamefont {Brecht}},
  \bibinfo {author} {\bibfnamefont {W.}~\bibnamefont {Mauerer}}, \ and\
  \bibinfo {author} {\bibfnamefont {C.}~\bibnamefont {Silberhorn}},\
  }\href@noop {} {\bibfield  {journal} {\bibinfo  {journal} {New J. Phys.}\
  }\textbf {\bibinfo {volume} {15}},\ \bibinfo {pages} {053038} (\bibinfo
  {year} {2013})}\BibitemShut {NoStop}%
\bibitem [{\citenamefont {Grice}\ and\ \citenamefont
  {Walmsley}(1997)}]{Grice:1997ht}%
  \BibitemOpen
  \bibfield  {author} {\bibinfo {author} {\bibfnamefont {W.~P.}\ \bibnamefont
  {Grice}}\ and\ \bibinfo {author} {\bibfnamefont {I.~A.}\ \bibnamefont
  {Walmsley}},\ }\href@noop {} {\bibfield  {journal} {\bibinfo  {journal}
  {Phys. Rev. A}\ }\textbf {\bibinfo {volume} {56}},\ \bibinfo {pages} {1627}
  (\bibinfo {year} {1997})}\BibitemShut {NoStop}%
\bibitem [{\citenamefont {Rubin}\ \emph {et~al.}(1994)\citenamefont {Rubin},
  \citenamefont {Klyshko}, \citenamefont {Shih},\ and\ \citenamefont
  {Sergienko}}]{Rubin:1994theory}%
  \BibitemOpen
  \bibfield  {author} {\bibinfo {author} {\bibfnamefont {M.~H.}\ \bibnamefont
  {Rubin}}, \bibinfo {author} {\bibfnamefont {D.~N.}\ \bibnamefont {Klyshko}},
  \bibinfo {author} {\bibfnamefont {Y.~H.}\ \bibnamefont {Shih}}, \ and\
  \bibinfo {author} {\bibfnamefont {A.~V.}\ \bibnamefont {Sergienko}},\
  }\href@noop {} {\bibfield  {journal} {\bibinfo  {journal} {Phys. Rev. A}\
  }\textbf {\bibinfo {volume} {50}},\ \bibinfo {pages} {5122} (\bibinfo {year}
  {1994})}\BibitemShut {NoStop}%
\bibitem [{\citenamefont {Saleh}\ \emph {et~al.}(1998)\citenamefont {Saleh},
  \citenamefont {Jost}, \citenamefont {Fei},\ and\ \citenamefont
  {Teich}}]{Saleh:1998vl}%
  \BibitemOpen
  \bibfield  {author} {\bibinfo {author} {\bibfnamefont {B.~E.~A.}\
  \bibnamefont {Saleh}}, \bibinfo {author} {\bibfnamefont {B.~M.}\ \bibnamefont
  {Jost}}, \bibinfo {author} {\bibfnamefont {H.-B.}\ \bibnamefont {Fei}}, \
  and\ \bibinfo {author} {\bibfnamefont {M.~C.}\ \bibnamefont {Teich}},\
  }\href@noop {} {\bibfield  {journal} {\bibinfo  {journal} {Phys. Rev. Lett.}\
  }\textbf {\bibinfo {volume} {80}},\ \bibinfo {pages} {3483} (\bibinfo {year}
  {1998})}\BibitemShut {NoStop}%
\bibitem [{\citenamefont {Dorfman}\ \emph {et~al.}(2016)\citenamefont
  {Dorfman}, \citenamefont {Schlawin},\ and\ \citenamefont
  {Mukamel}}]{Dorfman:2016da}%
  \BibitemOpen
  \bibfield  {author} {\bibinfo {author} {\bibfnamefont {K.~E.}\ \bibnamefont
  {Dorfman}}, \bibinfo {author} {\bibfnamefont {F.}~\bibnamefont {Schlawin}}, \
  and\ \bibinfo {author} {\bibfnamefont {S.}~\bibnamefont {Mukamel}},\
  }\href@noop {} {\bibfield  {journal} {\bibinfo  {journal} {Rev. Mod. Phys.}\
  }\textbf {\bibinfo {volume} {88}},\ \bibinfo {pages} {045008} (\bibinfo
  {year} {2016})}\BibitemShut {NoStop}%
\bibitem [{\citenamefont {Quesada}\ and\ \citenamefont
  {Bra{\'n}czyk}(2019)}]{Quesada:2019gj}%
  \BibitemOpen
  \bibfield  {author} {\bibinfo {author} {\bibfnamefont {N.}~\bibnamefont
  {Quesada}}\ and\ \bibinfo {author} {\bibfnamefont {A.~M.}\ \bibnamefont
  {Bra{\'n}czyk}},\ }\href@noop {} {\bibfield  {journal} {\bibinfo  {journal}
  {Phys. Rev. A}\ }\textbf {\bibinfo {volume} {99}},\ \bibinfo {pages} {013830}
  (\bibinfo {year} {2019})}\BibitemShut {NoStop}%
\bibitem [{\citenamefont {Grice}\ \emph {et~al.}(2001)\citenamefont {Grice},
  \citenamefont {U'Ren},\ and\ \citenamefont {Walmsley}}]{Grice:2001jc}%
  \BibitemOpen
  \bibfield  {author} {\bibinfo {author} {\bibfnamefont {W.~P.}\ \bibnamefont
  {Grice}}, \bibinfo {author} {\bibfnamefont {A.~B.}\ \bibnamefont {U'Ren}}, \
  and\ \bibinfo {author} {\bibfnamefont {I.~A.}\ \bibnamefont {Walmsley}},\
  }\href@noop {} {\bibfield  {journal} {\bibinfo  {journal} {Phys. Rev. A}\
  }\textbf {\bibinfo {volume} {64}},\ \bibinfo {pages} {063815} (\bibinfo
  {year} {2001})}\BibitemShut {NoStop}%
\bibitem [{\citenamefont {Lee}\ and\ \citenamefont
  {Goodson}(2006)}]{Lee:2006id}%
  \BibitemOpen
  \bibfield  {author} {\bibinfo {author} {\bibfnamefont {D.-I.}\ \bibnamefont
  {Lee}}\ and\ \bibinfo {author} {\bibfnamefont {T.}~\bibnamefont {Goodson}},\
  }\href@noop {} {\bibfield  {journal} {\bibinfo  {journal} {J. Phys. Chem. B}\
  }\textbf {\bibinfo {volume} {110}},\ \bibinfo {pages} {25582} (\bibinfo
  {year} {2006})}\BibitemShut {NoStop}%
\bibitem [{\citenamefont {MacLean}\ \emph {et~al.}(2018)\citenamefont
  {MacLean}, \citenamefont {Donohue},\ and\ \citenamefont
  {Resch}}]{maclean:2018direct}%
  \BibitemOpen
  \bibfield  {author} {\bibinfo {author} {\bibfnamefont {J.-P.~W.}\
  \bibnamefont {MacLean}}, \bibinfo {author} {\bibfnamefont {J.~M.}\
  \bibnamefont {Donohue}}, \ and\ \bibinfo {author} {\bibfnamefont {K.~J.}\
  \bibnamefont {Resch}},\ }\href@noop {} {\bibfield  {journal} {\bibinfo
  {journal} {Phys. Rev. Lett.}\ }\textbf {\bibinfo {volume} {120}},\ \bibinfo
  {pages} {053601} (\bibinfo {year} {2018})}\BibitemShut {NoStop}%
\bibitem [{\citenamefont {Loudon}(2000)}]{loudon2000quantum}%
  \BibitemOpen
  \bibfield  {author} {\bibinfo {author} {\bibfnamefont {R.}~\bibnamefont
  {Loudon}},\ }\href@noop {} {\emph {\bibinfo {title} {The Quantum Theory of
  Light}}},\ \bibinfo {edition} {3rd}\ ed.\ (\bibinfo  {publisher} {Oxford
  University Press},\ \bibinfo {address} {Oxford},\ \bibinfo {year}
  {2000})\BibitemShut {NoStop}%
\bibitem [{\citenamefont {Zueco}\ \emph {et~al.}(2009)\citenamefont {Zueco},
  \citenamefont {Reuther}, \citenamefont {Kohler},\ and\ \citenamefont
  {H{\"a}nggi}}]{Zueco:2009it}%
  \BibitemOpen
  \bibfield  {author} {\bibinfo {author} {\bibfnamefont {D.}~\bibnamefont
  {Zueco}}, \bibinfo {author} {\bibfnamefont {G.~M.}\ \bibnamefont {Reuther}},
  \bibinfo {author} {\bibfnamefont {S.}~\bibnamefont {Kohler}}, \ and\ \bibinfo
  {author} {\bibfnamefont {P.}~\bibnamefont {H{\"a}nggi}},\ }\href@noop {}
  {\bibfield  {journal} {\bibinfo  {journal} {Phys. Rev. A}\ }\textbf {\bibinfo
  {volume} {80}},\ \bibinfo {pages} {033846} (\bibinfo {year}
  {2009})}\BibitemShut {NoStop}%
\bibitem [{\citenamefont {Chenu}\ \emph {et~al.}(2014)\citenamefont {Chenu},
  \citenamefont {Mal{\'{y}}},\ and\ \citenamefont {Man{\v
  c}al}}]{Chenu:2014gr}%
  \BibitemOpen
  \bibfield  {author} {\bibinfo {author} {\bibfnamefont {A.}~\bibnamefont
  {Chenu}}, \bibinfo {author} {\bibfnamefont {P.}~\bibnamefont {Mal{\'{y}}}}, \
  and\ \bibinfo {author} {\bibfnamefont {T.}~\bibnamefont {Man{\v c}al}},\
  }\href@noop {} {\bibfield  {journal} {\bibinfo  {journal} {Chem. Phys.}\
  }\textbf {\bibinfo {volume} {439}},\ \bibinfo {pages} {100} (\bibinfo {year}
  {2014})}\BibitemShut {NoStop}%
\bibitem [{\citenamefont {Tscherbul}\ and\ \citenamefont
  {Brumer}(2014)}]{Tscherbul:2014cf}%
  \BibitemOpen
  \bibfield  {author} {\bibinfo {author} {\bibfnamefont {T.~V.}\ \bibnamefont
  {Tscherbul}}\ and\ \bibinfo {author} {\bibfnamefont {P.}~\bibnamefont
  {Brumer}},\ }\href@noop {} {\bibfield  {journal} {\bibinfo  {journal} {Phys.
  Rev. A}\ }\textbf {\bibinfo {volume} {89}},\ \bibinfo {pages} {013423}
  (\bibinfo {year} {2014})}\BibitemShut {NoStop}%
\bibitem [{\citenamefont {Pittman}\ \emph {et~al.}(1995)\citenamefont
  {Pittman}, \citenamefont {Shih}, \citenamefont {Strekalov},\ and\
  \citenamefont {Sergienko}}]{Pittman:1995un}%
  \BibitemOpen
  \bibfield  {author} {\bibinfo {author} {\bibfnamefont {T.~B.}\ \bibnamefont
  {Pittman}}, \bibinfo {author} {\bibfnamefont {Y.~H.}\ \bibnamefont {Shih}},
  \bibinfo {author} {\bibfnamefont {D.~V.}\ \bibnamefont {Strekalov}}, \ and\
  \bibinfo {author} {\bibfnamefont {A.~V.}\ \bibnamefont {Sergienko}},\
  }\href@noop {} {\bibfield  {journal} {\bibinfo  {journal} {Phys. Rev. A}\
  }\textbf {\bibinfo {volume} {52}},\ \bibinfo {pages} {R3429} (\bibinfo {year}
  {1995})}\BibitemShut {NoStop}%
\bibitem [{\citenamefont {Yabushita}\ and\ \citenamefont
  {Kobayashi}(2004)}]{Yabushita:2004hy}%
  \BibitemOpen
  \bibfield  {author} {\bibinfo {author} {\bibfnamefont {A.}~\bibnamefont
  {Yabushita}}\ and\ \bibinfo {author} {\bibfnamefont {T.}~\bibnamefont
  {Kobayashi}},\ }\href@noop {} {\bibfield  {journal} {\bibinfo  {journal}
  {Phys. Rev. A}\ }\textbf {\bibinfo {volume} {69}},\ \bibinfo {pages} {013806}
  (\bibinfo {year} {2004})}\BibitemShut {NoStop}%
\bibitem [{\citenamefont {Kalachev}\ \emph {et~al.}(2007)\citenamefont
  {Kalachev}, \citenamefont {Kalashnikov}, \citenamefont {Kalinkin},
  \citenamefont {Mitrofanova}, \citenamefont {Shkalikov},\ and\ \citenamefont
  {Samartsev}}]{Kalachev:2013kh}%
  \BibitemOpen
  \bibfield  {author} {\bibinfo {author} {\bibfnamefont {A.~A.}\ \bibnamefont
  {Kalachev}}, \bibinfo {author} {\bibfnamefont {D.~A.}\ \bibnamefont
  {Kalashnikov}}, \bibinfo {author} {\bibfnamefont {A.~A.}\ \bibnamefont
  {Kalinkin}}, \bibinfo {author} {\bibfnamefont {T.~G.}\ \bibnamefont
  {Mitrofanova}}, \bibinfo {author} {\bibfnamefont {A.~V.}\ \bibnamefont
  {Shkalikov}}, \ and\ \bibinfo {author} {\bibfnamefont {V.~V.}\ \bibnamefont
  {Samartsev}},\ }\href@noop {} {\bibfield  {journal} {\bibinfo  {journal}
  {Laser Phys. Lett.}\ }\textbf {\bibinfo {volume} {4}},\ \bibinfo {pages}
  {722} (\bibinfo {year} {2007})}\BibitemShut {NoStop}%
\bibitem [{\citenamefont {Raymer}\ \emph {et~al.}(2013)\citenamefont {Raymer},
  \citenamefont {Marcus}, \citenamefont {Widom},\ and\ \citenamefont
  {Vitullo}}]{Raymer:2013kj}%
  \BibitemOpen
  \bibfield  {author} {\bibinfo {author} {\bibfnamefont {M.~G.}\ \bibnamefont
  {Raymer}}, \bibinfo {author} {\bibfnamefont {A.~H.}\ \bibnamefont {Marcus}},
  \bibinfo {author} {\bibfnamefont {J.~R.}\ \bibnamefont {Widom}}, \ and\
  \bibinfo {author} {\bibfnamefont {D.~L.~P.}\ \bibnamefont {Vitullo}},\
  }\href@noop {} {\bibfield  {journal} {\bibinfo  {journal} {J. Phys. Chem. B}\
  }\textbf {\bibinfo {volume} {117}},\ \bibinfo {pages} {15559} (\bibinfo
  {year} {2013})}\BibitemShut {NoStop}%
\bibitem [{\citenamefont {Dorfman}\ \emph {et~al.}(2014)\citenamefont
  {Dorfman}, \citenamefont {Schlawin},\ and\ \citenamefont
  {Mukamel}}]{Dorfman:2014bn}%
  \BibitemOpen
  \bibfield  {author} {\bibinfo {author} {\bibfnamefont {K.~E.}\ \bibnamefont
  {Dorfman}}, \bibinfo {author} {\bibfnamefont {F.}~\bibnamefont {Schlawin}}, \
  and\ \bibinfo {author} {\bibfnamefont {S.}~\bibnamefont {Mukamel}},\
  }\href@noop {} {\bibfield  {journal} {\bibinfo  {journal} {J. Phys. Chem.
  Lett.}\ }\textbf {\bibinfo {volume} {5}},\ \bibinfo {pages} {2843} (\bibinfo
  {year} {2014})}\BibitemShut {NoStop}%
\bibitem [{\citenamefont {Schlawin}\ \emph {et~al.}(2016)\citenamefont
  {Schlawin}, \citenamefont {Dorfman},\ and\ \citenamefont
  {Mukamel}}]{Schlawin:2016er}%
  \BibitemOpen
  \bibfield  {author} {\bibinfo {author} {\bibfnamefont {F.}~\bibnamefont
  {Schlawin}}, \bibinfo {author} {\bibfnamefont {K.~E.}\ \bibnamefont
  {Dorfman}}, \ and\ \bibinfo {author} {\bibfnamefont {S.}~\bibnamefont
  {Mukamel}},\ }\href@noop {} {\bibfield  {journal} {\bibinfo  {journal} {Phys.
  Rev. A}\ }\textbf {\bibinfo {volume} {93}},\ \bibinfo {pages} {023807}
  (\bibinfo {year} {2016})}\BibitemShut {NoStop}%
\bibitem [{\citenamefont {Du}(2015)}]{Du:2015gc}%
  \BibitemOpen
  \bibfield  {author} {\bibinfo {author} {\bibfnamefont {S.}~\bibnamefont
  {Du}},\ }\href@noop {} {\bibfield  {journal} {\bibinfo  {journal} {Phys. Rev.
  A}\ }\textbf {\bibinfo {volume} {92}},\ \bibinfo {pages} {043836} (\bibinfo
  {year} {2015})}\BibitemShut {NoStop}%
\bibitem [{\citenamefont {Glauber}(2007)}]{Glauber2007book}%
  \BibitemOpen
  \bibfield  {author} {\bibinfo {author} {\bibfnamefont {R.~J.}\ \bibnamefont
  {Glauber}},\ }\href@noop {} {\emph {\bibinfo {title} {Quantum Theory of
  Optical Coherence: Selected Papers and Lectures}}}\ (\bibinfo  {publisher}
  {Wiley-VCH},\ \bibinfo {address} {Berlin},\ \bibinfo {year}
  {2007})\BibitemShut {NoStop}%
\bibitem [{Note1()}]{Note1}%
  \BibitemOpen
  \bibinfo {note} {The correlation function in Eq.~\protect \textup {\hbox
  {\mathsurround \z@ \protect \normalfont (\ignorespaces \ref
  {eq:averaged-RDM}\unskip \@@italiccorr )}} is practically identical to the
  first-order temporal correlation function for individual realizations of the
  transform-limited classical laser pulses investigated by Chenu and Brumer to
  yield the thermal light result \cite {Chenu:2016cx} when $\protect
  \mathaccentV {bar}016{n}(\omega )$ in Eq.~\protect \textup {\hbox
  {\mathsurround \z@ \protect \normalfont (\ignorespaces \ref
  {eq:approximated-pseudo-field}\unskip \@@italiccorr )}} is replaced with
  $\protect \mathaccentV {bar}016{n}_{\protect \rm th}(\omega )$.}\BibitemShut
  {Stop}%
\bibitem [{\citenamefont {Jin}\ \emph {et~al.}(2011)\citenamefont {Jin},
  \citenamefont {Zhang}, \citenamefont {Shimizu}, \citenamefont {Matsuda},
  \citenamefont {Mitsumori}, \citenamefont {Kosaka},\ and\ \citenamefont
  {Edamatsu}}]{Jin:2011dt}%
  \BibitemOpen
  \bibfield  {author} {\bibinfo {author} {\bibfnamefont {R.-B.}\ \bibnamefont
  {Jin}}, \bibinfo {author} {\bibfnamefont {J.}~\bibnamefont {Zhang}}, \bibinfo
  {author} {\bibfnamefont {R.}~\bibnamefont {Shimizu}}, \bibinfo {author}
  {\bibfnamefont {N.}~\bibnamefont {Matsuda}}, \bibinfo {author} {\bibfnamefont
  {Y.}~\bibnamefont {Mitsumori}}, \bibinfo {author} {\bibfnamefont
  {H.}~\bibnamefont {Kosaka}}, \ and\ \bibinfo {author} {\bibfnamefont
  {K.}~\bibnamefont {Edamatsu}},\ }\href@noop {} {\bibfield  {journal}
  {\bibinfo  {journal} {Phys. Rev. A}\ }\textbf {\bibinfo {volume} {83}},\
  \bibinfo {pages} {031805(R)} (\bibinfo {year} {2011})}\BibitemShut {NoStop}%
\bibitem [{Note2()}]{Note2}%
  \BibitemOpen
  \bibinfo {note} {After submitting this paper, we became aware of Ref.~\cite
  {Deng:2019cu}, in which C.-Y. Lu, J.-W. Pan and coworkers experimentally
  demonstrated nonclassical interference between sunlight and single photons
  from a quantum dot. For measuring the Hong-Ou-Mandel interference, only the
  single-photon events of the sunlight were registered.}\BibitemShut {Stop}%
\bibitem [{\citenamefont {Mukamel}(1995)}]{Mukamel:1995us}%
  \BibitemOpen
  \bibfield  {author} {\bibinfo {author} {\bibfnamefont {S.}~\bibnamefont
  {Mukamel}},\ }\href@noop {} {\emph {\bibinfo {title} {{Principles of
  Nonlinear Optical Spectroscopy}}}}\ (\bibinfo  {publisher} {Oxford University
  Press},\ \bibinfo {address} {New York},\ \bibinfo {year} {1995})\BibitemShut
  {NoStop}%
\bibitem [{\citenamefont {Deng}\ \emph {et~al.}(2019)\citenamefont {Deng},
  \citenamefont {Wang}, \citenamefont {Ding}, \citenamefont {Duan},
  \citenamefont {Qin}, \citenamefont {Chen}, \citenamefont {He}, \citenamefont
  {He}, \citenamefont {Li}, \citenamefont {Li}, \citenamefont {Peng},
  \citenamefont {Matekole}, \citenamefont {Byrnes}, \citenamefont {Schneider},
  \citenamefont {Kamp}, \citenamefont {Wang}, \citenamefont {Dowling},
  \citenamefont {H{\"o}fling}, \citenamefont {Lu}, \citenamefont {Scully},\
  and\ \citenamefont {Pan}}]{Deng:2019cu}%
  \BibitemOpen
  \bibfield  {author} {\bibinfo {author} {\bibfnamefont {Y.-H.}\ \bibnamefont
  {Deng}}, \bibinfo {author} {\bibfnamefont {H.}~\bibnamefont {Wang}}, \bibinfo
  {author} {\bibfnamefont {X.}~\bibnamefont {Ding}}, \bibinfo {author}
  {\bibfnamefont {Z.~C.}\ \bibnamefont {Duan}}, \bibinfo {author}
  {\bibfnamefont {J.}~\bibnamefont {Qin}}, \bibinfo {author} {\bibfnamefont
  {M.~C.}\ \bibnamefont {Chen}}, \bibinfo {author} {\bibfnamefont
  {Y.}~\bibnamefont {He}}, \bibinfo {author} {\bibfnamefont {Y.-M.}\
  \bibnamefont {He}}, \bibinfo {author} {\bibfnamefont {J.-P.}\ \bibnamefont
  {Li}}, \bibinfo {author} {\bibfnamefont {Y.-H.}\ \bibnamefont {Li}}, \bibinfo
  {author} {\bibfnamefont {L.-C.}\ \bibnamefont {Peng}}, \bibinfo {author}
  {\bibfnamefont {E.~S.}\ \bibnamefont {Matekole}}, \bibinfo {author}
  {\bibfnamefont {T.}~\bibnamefont {Byrnes}}, \bibinfo {author} {\bibfnamefont
  {C.}~\bibnamefont {Schneider}}, \bibinfo {author} {\bibfnamefont
  {M.}~\bibnamefont {Kamp}}, \bibinfo {author} {\bibfnamefont {D.-W.}\
  \bibnamefont {Wang}}, \bibinfo {author} {\bibfnamefont {J.~P.}\ \bibnamefont
  {Dowling}}, \bibinfo {author} {\bibfnamefont {S.}~\bibnamefont
  {H{\"o}fling}}, \bibinfo {author} {\bibfnamefont {C.-Y.}\ \bibnamefont {Lu}},
  \bibinfo {author} {\bibfnamefont {M.~O.}\ \bibnamefont {Scully}}, \ and\
  \bibinfo {author} {\bibfnamefont {J.-W.}\ \bibnamefont {Pan}},\ }\href@noop
  {} {\bibfield  {journal} {\bibinfo  {journal} {Phys. Rev. Lett.}\ }\textbf
  {\bibinfo {volume} {123}},\ \bibinfo {pages} {080401} (\bibinfo {year}
  {2019})}\BibitemShut {NoStop}%
\end{thebibliography}

%

\end{document}